\documentclass[aip,apl,twocolumn,superscriptaddress,numerical]{revtex4}
\usepackage{amssymb,amsmath}
\usepackage{graphicx}
\usepackage{dcolumn}
\usepackage{multirow}
\usepackage{color}
\usepackage{hyperref}
\usepackage{epsfig}
\usepackage{bm}
\usepackage{placeins}
\usepackage{tabularx}

\newcommand{\unit}[1]{~\mathrm{#1}}

\begin{document}

\title{\textbf{\fontfamily{phv}\selectfont Reconfigurable quadruple quantum dots in a silicon nanowire transistor}}
\author{A.~C.~Betz}
\email{ab2106@cam.ac.uk}
\affiliation{Hitachi Cambridge Laboratory, J. J. Thomson Avenue, Cambridge CB3 0HE, United Kingdom}
\author{M.~L.~V.~Tagliaferri}
\affiliation{Laboratorio MDM, CNR-IMM, Via C. Olivetti 2, 20864 Agrate Brianza (MB), Italy}
\affiliation{Dipartimento di Scienza dei Materiali, Università di Milano-Bicocca, Via Cozzi 53, 20125 Milano, Italy}
\author{M.~Vinet}
\affiliation{CEA/LETI-MINATEC, CEA-Grenoble, 17 rue des martyrs, F-38054 Grenoble, France}
\author{M. Brostr\"{o}m}
\affiliation{Hitachi Cambridge Laboratory, J. J. Thomson Avenue, Cambridge CB3 0HE, United Kingdom}
\author{M.~Sanquer}
\affiliation{SPSMS, UMR-E CEA / UJF-Grenoble 1, INAC, 17 rue des Martyrs, 38054 Grenoble, France}
\author{A.~J.~Ferguson}
\affiliation{Cavendish Laboratory, University of Cambridge, Cambridge CB3 0HE, United Kingdom}
\author{M.~F.~Gonzalez-Zalba}
\affiliation{Hitachi Cambridge Laboratory, J. J. Thomson Avenue, Cambridge CB3 0HE, United Kingdom}

\date{\today}

\begin{abstract}
We present a novel reconfigurable metal-oxide-semiconductor multi-gate transistor that can host a quadruple quantum dot in silicon. The device consist of an industrial quadruple-gate silicon nanowire field-effect transistor. Exploiting the corner effect, we study the versatility of the structure in the single quantum dot and the serial double quantum dot regimes and extract the relevant capacitance parameters. We address the fabrication variability of the quadruple-gate approach which, paired with improved silicon fabrication techniques, makes the corner state quantum dot approach a promising candidate for a scalable quantum information architecture. 
\end{abstract}

\maketitle

Semiconductor quantum bits relying on the charge or spin degree of freedom of a single electron, bound to a quantum dot (QD) or impurity atom, are considered promising candidates for the base elements of solid state quantum computing architectures \cite{ZwanenburgRevModPhys2013}. Building a successful quantum computer, however, requires a scalable multi-qubit approach to implement the necessary algorithms \cite{DiVincenzo2000}. Electron spins bound to silicon QDs are seen as promising candidates for this due to their long coherence time, electrical tunability and flexible coupling geometries \cite{Veldhorst2014,Kim2014,VeldhorstNature2015}. 
A further advantage of using Si is the possibility to integrate with current complementary-metal-oxide-semiconductor (CMOS) technology \cite{Angus2007,VeldhorstNature2015,Pla2013,CrippaPRB2015} and leverage its established industrial platform for large scale circuits.
Recently, the integration of Si quantum dots and double quantum dots (DQD) into CMOS technology has been taken a step further with reports of few-electron QDs, DQDs, and donor-QD hybrids created within industry-standard Si nanowire transistors \cite{Voisin2014NL,BetzNL2015,Gonzalez2015NC,UrdampilletaPRX2015}. Combined with a gate-based readout scheme that alleviates the need for a separate charge sensor \cite{Colless2013PRL,Gonzalez2015NC,BetzNL2015,UrdampilletaPRX2015,GonzalezNL2016} these approaches pave the way towards a large scale quantum computing architecture based on current CMOS technology.

In this Letter, we report on a reconfigurable QD and DQD system in a quadruple-gate CMOS transistor. It incorporates one, or a pair, of CMOS corner state quantum dots \cite{BetzNL2015,Gonzalez2015NC} in a variety of configurations.
\begin{figure}[htb]
\centering
\includegraphics[width=\columnwidth]{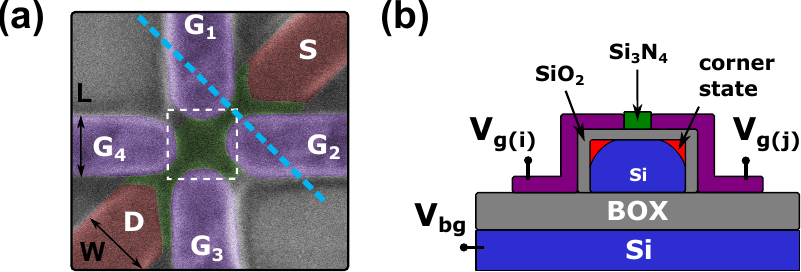}
\caption{Device geometry (a) Top-view scanning electron micrograph of the device. The Si$_3$N$_4$ region is highlighted in green, highly doped leads are red, and gate electrodes are coloured purple. (b) Sketch of device cross-section perpendicular to transport direction. A corner state quantum dot can be created under each gate at the top-most corners of the channel (see (a)). Each QD is controlled by its respective gate voltage $V_{g(x)}$, while a global backgate voltage $V_{bg}$ may be applied through the Si substrate. }\label{Fig1}
\end{figure}
\begin{figure}[thb]
\centering
\includegraphics[width=\columnwidth]{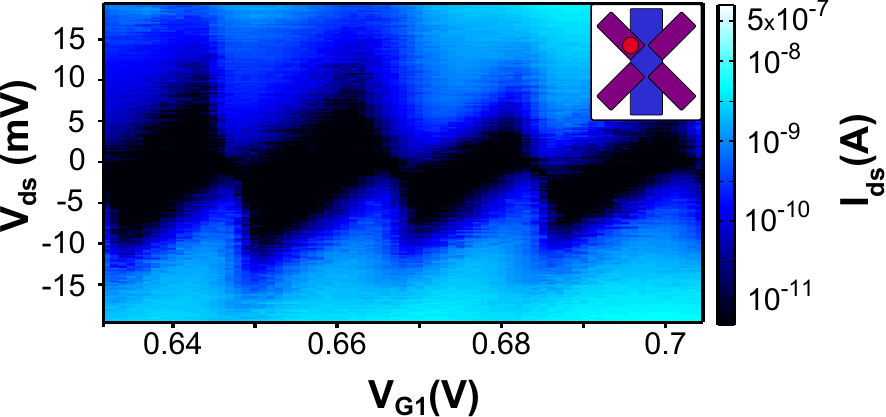}
\caption{Charge stability diagram of quantum dot under gate $G_1$ at $V_{g4}=1.4\unit{V}$, and $V_{g2}=V_{g3}=0\unit{V}$.}\label{Fig2}
\end{figure}
Each of the four gates can host an independently tunable quantum dot created in the square channel by electrostatic enhancement and confinement resulting from the top-gate electrodes and accompanying silicon nitride spacers. We characterise one exemplary single QD and demonstrate that different DQD configurations can be set at will. Building on previous demonstrations \cite{BetzNL2015,Gonzalez2015NC,UrdampilletaPRX2015}, our results provide a way to scale up CMOS quantum information architectures and to create reconfigurable silicon multi-dot arrangements.
\begin{figure*}[hbt]
\centering
\includegraphics[width=\textwidth]{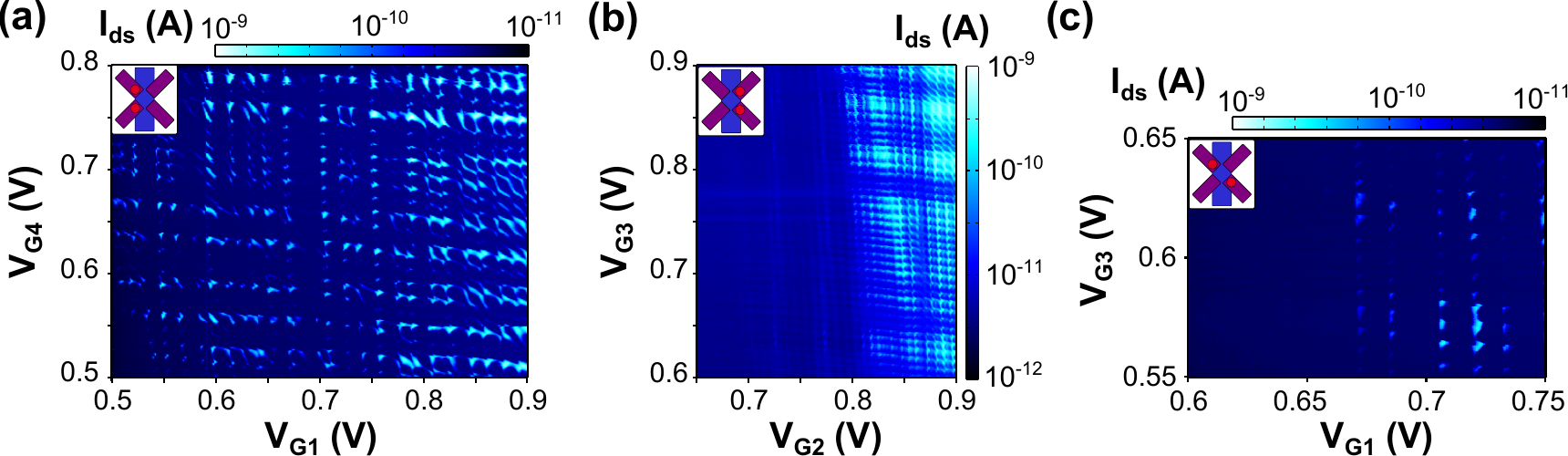}
\caption{Different DQD configurations established in the FDSOI nanowire transistor. Serial DQD created along gates (a) $G_1$-$G_4$, (b) $G_2$-$G_3$, and (c) $G_1$-$G_3$.}\label{Fig3}
\end{figure*}

The device presented here is a fully depleted silicon-on-insulator (FDSOI) nanowire field-effect transistor with four independently addressable top-gates. The poly-silicon top-gates of length $L=40\unit{nm}$ are arranged around the sides of a $82\unit{nm} \times 82\unit{nm}$ square at $45^{\circ}$ with respect to transport direction as shown in white dashes in the scanning electron micrograph of a similar device in Fig.\ref{Fig1}(a). Gate-to-gate distances are $S_{G1-G2}=S_{G2-G3}=30\unit{nm}$, and the channel width $W$ is $87\unit{nm}$. Fig.\ref{Fig1}(a) shows the Si$_3$N$_4$ spacers deposited between the gates and extending $40\unit{nm}$ towards source and drain in green. The channel below remains undoped generating the source, drain and inter-dot tunnel barriers.
Fig.\ref{Fig1}(b) shows a cross-section sketch of the device perpendicular to the direction of transport, taken at the line indicated by the cyan line in Fig.\ref{Fig1}(a). The device comprises an Si back plane serving as global backgate, topped by a $150\unit{nm}$ SiO$_2$ buried oxide. This is followed by the undoped, square Si (001) channel of thickness $12\unit{nm}$, which is achieved by etching down the SOI substrate prior to gate stack deposition.   
Each top-gate wraps around two faces of the intrinsic channel and is separated from the other top-gates by the Si$_3$N$_4$ spacers and from the channel by $5\unit{nm}$ of SiO$_2$ \cite{Hofheinz2006APL}. A quantum dot can be created under each gate at the top-most corners of the channel due to the so-called corner effect \cite{Sellier2007APL,Gonzalez2015NC,BetzNL2015}. Each QD is controlled by its respective gate voltage $V_{g(x)}$, while a global backgate voltage $V_{bg}$ may be applied through the Si substrate. All measurements shown in this Letter are in direct transport and unless otherwise stated were taken at $V_{bg}=5\unit{V}$, to enhance the dot-to-dot and dot-to-lead couplings \cite{Roche2012APL}.

Corner state quantum dots in Si nanowire transistors have been reported for different single and double gate topologies \cite{BetzNL2015,Gonzalez2015NC,Voisin2014NL} and using both quantum dots and dopants \cite{UrdampilletaPRX2015}. In our novel quadruple gate configuration, we first of all confirm the creation of a single QD under a top-gate using gate $G_1$ as an example. In order to allow for a source-drain current, $G_4$ is pulled high above threshold to $V_{g4}=1.4\unit{V}$, while the gates on the opposite channel side remain well below threshold at $V_{g2}=V_{g3}=0\unit{V}$. The backgate voltage is set to $0\unit{V}$. This provides a single QD under gate $G_1$, as can be seen from the stability map of Fig.\ref{Fig2}. We extract a charging energy $E_{c,G1}\simeq 9\unit{meV}$ and source and drain capacitances $C_{s,G1}\simeq 2\unit{aF}$ and $C_{d,G1}\simeq 1\unit{aF}$ from the slope of the Coulomb diamonds and the gate voltage period.
%Both the $V_{g1}$ Coulomb period and the charging energy vary indicating that the QD is close to the few electron regime \cite{TaruchaPRL1993,Kouwenhoven1997}. 
Single QDs embedded in a multi-gate structure like the one investigated here could host single electron spin qubits, or be used as single-electron transistors for charge detection of a nearby dot.
%could serve as a basis for single electron spin qubits, interacting with other single spin qubits or qubits based on double quantum dots. %\textcolor{red}{get better plot here, e.g. in multi-el regime, plus a map for G2}

\begin{figure}[htb]
\centering
\includegraphics[width=\columnwidth]{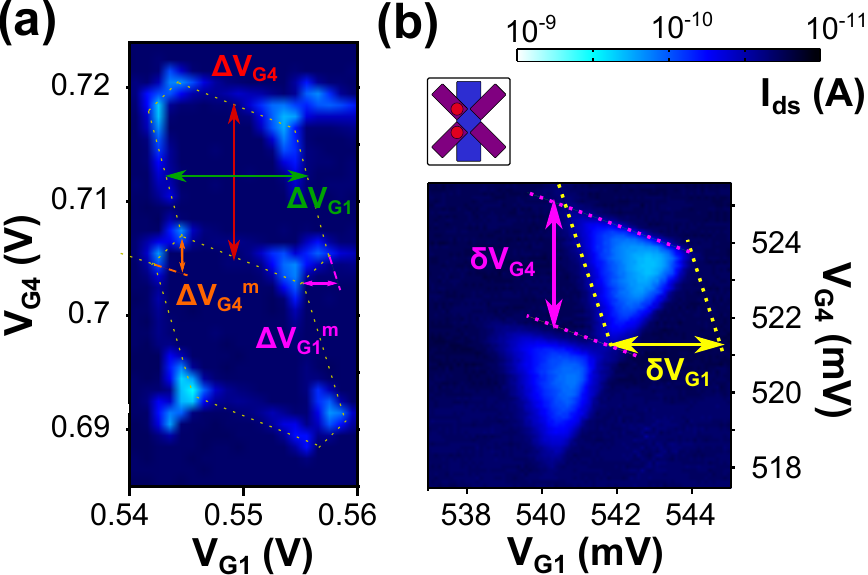}
\caption{Zoom of $G_1$-$G_4$ honeycomb diagram. (a) Honeycomb diagram along several bias triangles, outlined by yellow dashed lines. (b) Bias triangles at voltages below (a). }\label{Fig4}
\end{figure}

The ability to create multi-QD configurations is, as mentioned above, an important ingredient in the pursuit of a scalable semiconductor QIP architecture. In the following paragraphs we demonstrate that different configuration of serial DQDs can be formed in the nanowire transistor and extract the DQD coupling capacitances. All following measurements were carried out at a small source-drain voltage $V_{ds}=2\unit{mV}$.
First of all, we configure the device as a serial DQD on the $G_1$-$G_4$ axis, by setting $V_{g2}=V_{g3}=0\unit{V}$. Fig.\ref{Fig3}(a) shows the resulting DC current $I_{ds}$ as a function of $V_{g1}$ and $V_{g4}$. In the stability map, we observe a honeycomb pattern and at its vertices we find the so-called bias triangles, i.e. periodic regions of enhanced current resulting from the alignment of both QD energy levels within the bias window, indicative of a serial DQD \cite{Wiel2002}. At low voltages on both $G_1$ and $G_4$ the stability map shows the checker box behaviour characteristic for low inter-dot coupling, transitioning to the aforementioned honeycomb arrangement for intermediate coupling strength. At elevated top-gate voltages  the inter-dot coupling increases, resulting in undulated diagonal lines of enhanced conductance indicating that the QDs are strongly tunnel coupled \cite{Lai2011}. The voltage spacing of Coulomb oscillations in the low to intermediate coupling regime is $\Delta V_{g1}\simeq 11\unit{mV}$ and $\Delta V_{g4}\simeq 14\unit{mV}$ (see also Fig.\ref{Fig4}(a)). Both spacings are very regular across the voltage range studied here, indicating stable QD position and size. 
Fig.\ref{Fig3}(b) presents the opposite configuration to Fig.\ref{Fig3}(a), a serial DQD along $G_2$-$G_3$ with now $V_{g1}=V_{g4}=0\unit{V}$. Similarly to Fig.\ref{Fig3}(a) we obtain a honeycomb diagram, albeit a more compressed one with $\Delta V_{g2}\simeq 9\unit{mV}$ and $\Delta V_{g3}\simeq 8\unit{mV}$, indicating larger QDs.
Finally, Fig.\ref{Fig3}(c) shows the stability map for a diagonal DQD using $G_1$ and $G_3$ as constituent quantum dots. Here, the overall current is reduced compared to the previous two configurations, which we attribute to the larger distance between the dots under $G_2$ and $G_3$ and hence reduced iter-dot tunnel rates. We extract Coulomb spacings $\Delta V_{g1}\simeq 13\unit{mV}$ and $\Delta V_{g3}\simeq 7\unit{mV}$, approximately in line with the previous results. 
%In both Fig.\ref{Fig3}(b) and (c) a third modulation is present varying the total current levels. This could be due to interaction . 
%The quadruple gate transistor can also be brought into parallel DQD arrangements similar to the one reported in Ref.\cite{BetzNL2015}. %This requires however for the second parallel pair of gates to be pulled high above threshold, obscuring results due to gate leakage mainly from gates $G_2$ and $G_3$ at gate voltages exceeding $1\unit{V}$.
Similar results have been obtained for the DQD along the $G_2$-$G_4$ axis (not shown here).

Last but not least, we study the serial DQD along $G_1$-$G_4$ in more detail to obtain estimates of the capacitances involved in the system. From the honeycomb diagram outlined in Fig.\ref{Fig4}(a) we extract voltage spacings $\Delta V_{G1}\simeq 11\unit{mV}$ and $\Delta V_{G4}\simeq 14\unit{mV}$ between Coulomb oscillations. These spacings are linked to the gate capacitances of the two serial QDs as $\Delta V_{Gi} = e/C_{Gi}$ \cite{Wiel2002} and we thus estimate $C_{G1}\simeq 15\unit{aF}$ and $C_{G4}\simeq 12\unit{aF}$. Applying this analysis to the $G_2$-$G_3$ configuration, we find gate capacitances $C_{G2}\simeq 12\unit{aF}$ and $C_{G3}\simeq 20\unit{aF}$. The voltage spacings $\delta V_{G1}$ and $\delta V_{G4}$ extracted from Fig.\ref{Fig4}(b) in combination with the source-drain voltage $V_{sd}=2\unit{mV}$ provide us with an estimate for the lever arms $\alpha_{1(4)}$ and total capacitances $C_{1(4)}$ as $\alpha_{1(4)} \delta V_{G1(4)} = e|V_{sd}|=e\delta V_{G1(4)} C_{G1(4)}/C_{1(4)}$ \cite{Wiel2002}. We find lever arms $\alpha_{1(4)}\simeq 0.55\,(0.49)$ and total capacitances $C_{1(4)}\simeq 26\,(23)\unit{aF}$. These values in addition to $\Delta V_{g1(4)}^m$ from Fig.\ref{Fig4}(a) allow us now to also approximate the mutual capacitance linking the QDs under gates $G_1$ and $G_4$: $C_{m,1-4} = C_4 \Delta V_{g1}^m/\Delta V_{g1}\simeq 6\unit{aF}$. The same analysis carried out for the diagonal DQD along $G_1$-$G_3$ (not shown here) yields a mutual capacitance $C_{m,1-3}\simeq 4\unit{aF}$. %Table \ref{Tab1} summarises the extracted capacitance parameters.

Variability between nominally identical devices from different batches but also within a single batch fabricated at the same time is an important benchmarking parameter in CMOS technology. Comparing all four gate capacitances of the device presented here, we find a mean gate capacitance of about $16\unit{aF}$ and a variance of $25\%$. We attribute this variance to variations in gate dielectric thickness and quality, as well as gate misalignment relative to the channel. All three factors directly influence the enhancement mode corner state QDs. 
%The device examined here was not optimised for operation as DQD. 
%Taking into account the continuous improvement of CMOS fabrication techniques, which now e.g. routinely include high quality high-k dielectrics, corner state QDs and DQDs present a valid and scalable option as base elements of a silicon quantum information architecture.
%\begin{table}[htb]
%\centering
%\begin{tabularx}{\columnwidth}{ |X X | X  X  X  X  X  X | }
%  \hline		
%   & & 1 & 2 & 3 & 4 & 1-4 & 2-3 \\ \hline
%  $C_{Gi}$ & $\unit{(aF)}$ & 15 & 12 & 20 & 12 &  & \\
%  $C_{i}$ & $\unit{(aF)}$ & 26 &  &  & 23 & &  \\
%  $C_{i-j}^m$ & $\unit{(aF)}$ &  &  &  &  & 6 & 4 \\
%  \hline  
%\end{tabularx}
%\caption{Summary of capacitance analysis.}\label{Tab1}
%\end{table}

In conclusion, we have demonstrated a reconfigurable quantum dot and double quantum dot system based on a quadruple-gate CMOS transistor. 
%All four gates top-gates can be used to create a corner state QD and different configurations of DQDs may be set by choosing the appropriate gate voltages. 
Taking into account the continuous improvement of CMOS fabrication techniques and the single electron control demonstrated here, corner state quantum dots present a valid and scalable option as base elements of a silicon quantum information architecture.
Moreover, this approach can readily be extended to a larger number of gates providing a 1-D line of QDs and DQDs,%. qubits, e.g. singlet-triplet spin qubits, although single spin or other approaches are viable too. This approach and can be kept 
while remaining compact by further integrating high-sensitivity gate-based reflectometry readout \cite{Gonzalez2015NC,BetzNL2015}. The versatility of the corner state QD structure allows to use the gate-based sensing either for in-situ readout, but also for conventional charge sensing using an rf-single electron box for detection. Further experiments that can be envisioned with this architecture are e.g. electron charge and electron spin buses, measurements of long distance coherent coupling \cite{Braakman2013}, and single spin CCD \cite{Baart2015Nature}. 
\newline

The authors thank D.A. Williams and M. Fanciulli for support and discussion. The research leading to these results has been supported by the European Community's seventh Framework under the Grant Agreement No. 214989. The samples presented in this work were designed and fabricated by the AFSID project partners (www.afsid.eu). A.J. Ferguson acknowledges support from EPSRC (EP/K027018/1) and from his Hitachi research fellowship. M.L.V. Tagliaferri acknowledges support from the Short Term Mobility Program 2015 of Consiglio Nazionale delle Ricerche (CNR), Italy.
%\FloatBarrier
%\bibliographystyle{aipnum4-1}
%\bibliography{Bibliography}

%

\end{document}